\documentclass[manuscript]{aastex}
\usepackage{amsmath}
\usepackage[dvipdfm]{color}
\begin{document}

\title{A Theoretical Model of Pinching Current Sheet in Low-$\beta$ Plasmas}

\author{Satoshi TAKESHIGE\altaffilmark{1,2}, Shinsuke TAKASAO\altaffilmark{1,2} and Kazunari SHIBATA\altaffilmark{1}}
\email{takeshige@kwasan.kyoto-u.ac.jp}

\altaffiltext{1}{Kwasan and Hida Observatories, Kyoto University, 17 Ohmine-cho, Kita Kwazan, Yamashina, Kyoto 607-8471}
\altaffiltext{2}{Department of Astronomy, Kyoto University, Kitashirakawa-Oiwake-cho, Sakyo-ku, Kyoto 606-8502}

\begin{abstract}
Magnetic reconnection is an important physical process in various explosive phenomena in the universe.
In the previous studies, it was found that fast reconnection takes place when the thickness of a current sheet becomes on the order of a microscopic  length such as the ion larmor radius or the ion inertial length.
In this study, we investigated the pinching process of a current sheet by the Lorentz force in a low-$\beta$ plasma using one-dimensional magnetohydrodynamics (MHD) simulations.
It is known that there is an exact self-similar solution for this problem that neglects gas pressure.
We compared the non-linear MHD dynamics with the analytic self-similar solution.
From the MHD simulations, we found that with the gas pressure included the implosion process deviates from the analytic self-similar solution as $t \rightarrow t_0$, where $t_0$ is the explosion time when the thickness of a current sheet of the analytic solution becomes 0.
We also found a pair of MHD fast-mode shocks are generated and propagate after the formation of the pinched current sheet as $t \rightarrow t_0$.
On the basis of the Rankine-Hugoniot relations, we derived the scaling law of the physical quantities with respect to the initial plasma beta in the pinched current sheet.
Our study could help us to estimate the physical quantities in the pinched current sheet formed in a low-$\beta$ plasma.
\end{abstract}

\keywords{Magnetohydrodynamics (MHD) - Stars:sun - Stars: magnetic field}

\section{Introduction}
It has been known that magnetic reconnection plays important roles in many energetic explosions like solar flares \citep{zweibel09,shibata11}.
The knowledge about solar flares has been applied to other explosive phenomena, like superflares on solar type stars \citep{maehara12} and soft gamma repeaters \citep{lyutikov06,masada10,meng14}.
In a classical standard model of solar flares, the magnetic energy is rapidly converted to the thermal and kinetic energies of plasmas through magnetic reconnection \citep{carmichael64,sturrock66,hirayama74,kopp76,shibata95,shibata11}.
Such rapid reconnection requires a much shorter current sheet width than its length \citep{yamada10}.
Therefore, how thin current sheets are established is a central issue for understanding the origin of explosive phenomena.

Magnetohydrodynamics (MHD) processes to form a thin current sheet have been theoretically studied by many authors.
In \citet{forbes95} and \citet{lin00}, they numerically and analytically investigated the formation of a current sheet below an erupting flux rope during a solar flare.
There are many observational supports of the formation and the evolution of a current sheet between a cusp-shaped flaring loop and an erupting flux rope \citep{liu10}. 
The formation of a current sheet between an emerging flux and a pre-existing magnetic field in the solar corona has been numerically studied \citep{forbes84,shibata92,yokoyama96}.
In \citet{mclaughlin09}, the collapse of a null point to a current sheet triggered by MHD fast-mode waves was investigated.

A long and thin current sheet can form many fine scale current sheets inside it through some MHD instabilities.
Such a current sheet can be fragmented to form magnetic islands or plasmoids through the tearing or plasmoid instabilities \citep{furth63,loureiro07}.
The plasmoids are pulled against each other by a Lorentz force.
When plasmoids collide with each other, new current sheets will be formed between them.
This process can operate at multi-spatial scale, depending on the Lundquist number \citep{tajima87,shibata01,barta11}.
Observational support from the direct imaging observation of the plasmoid formation in a solar flare was given by \citet{takasao12} for the first time.
They also found that the plasma blobs (possibly plasmoids) in the current sheet collided or merged with each other before they were ejected from the current sheet.
The pinching process of a current sheet by a Lorentz force like the last two processes above is the main subject of this paper.

In this study, we investigated the implosion process in which current sheets are pinched by the magnetic pressure gradient.
In a previous study, the self-similar solution in the implosion process was analytically derived under the assumption that the gas pressure can be ignored  \citep{imshennik67, bulanov84, tajima87}.
In \citet{forbes82}, they investigated the implosion process including the gas pressure, and found the generation of two MHD fast-mode shocks as a result of a significant increase in the gas pressure at the center of the current sheet.
However, in their simulation, the current sheet was incompletely pinched because of the low spatial resolution.
To study the pinching process of the thin current sheet formed behind the shocks in detail, we performed a series of numerical simulations with a sufficient spatial resolution. 

In $\S 2$, the self-similar solution proposed by \citet{tajima87} will be introduced.
In $\S 3$, we will show the basic equations and initial conditions of the numerical simulations.
In $\S 4$, we will show the results of the numerical simulations and the power law scalings of the physical quantities in the pinched current sheets.
In $\S 5$, we will analytically derive the power law relation shown in $\S 4$.
Finally, we will summarize our conclusion in $\S 5$.

\section{Self-Similar Solution}
\begin{figure}
\begin{center}
\includegraphics[scale=1.00]{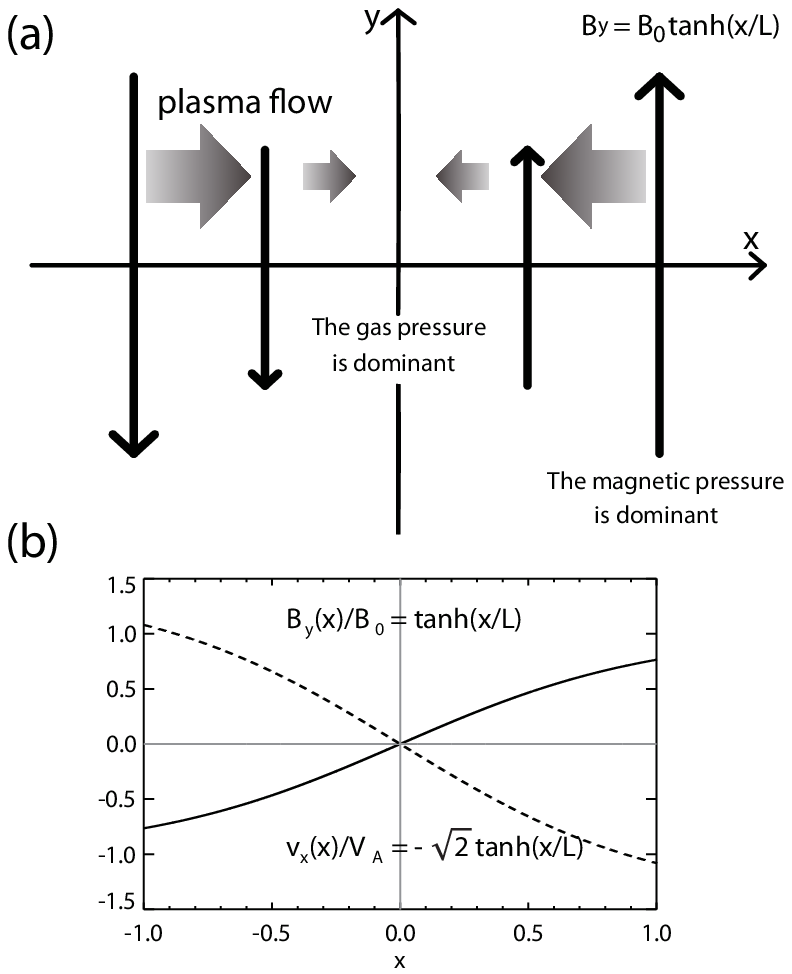}
\caption{(a) A schematic picture of the initial condition of our simulations.
The vertical solid arrows indicate magnetic field lines and the horizontal arrows indicate the plasma flow.
(b) The initial condition of a typical case of our simulations.
The solid line represents the initial distribution of $B_y$ and the dashed line represents the initial distribution of $v_x$.
$V_A$ represents the Alfven speed.
}
\label{fig11}
\end{center}
\end{figure}

The situation that we consider is similar to that of \citet{tajima87} where a current sheet is pinched by a Lorentz force (see Figure~\ref{fig11}).
They derived a self-similar solution using the two-fluid equations and Maxwell's equations.
Here we derive the same self-similar solution from 1D MHD equations to review the basic physics and assumptions of the solution.
We start from the basic 1D MHD equations:
\begin{align}
\frac{\partial}{\partial t} \rho & =  -\frac{\partial}{\partial x}\left( \rho v_x \right) , \label{57} \\
\frac{\partial}{\partial t} \left( \rho v_x\right) & =  -\frac{\partial}{\partial x} \left( p+\rho v^2\right) -\frac{j_z B_y}{c}, \label{58}\\
\frac{\partial}{\partial t} B_y & =  -\frac{\partial}{\partial x}\left( v_x B_y \right),\label{59}\\
j_z & =  \frac{c}{4\pi}\frac{\partial}{\partial x} B_y. 
\end{align}
Now we introduce dimensionless physical quantities as below:
\begin{align}
x ^\ast & \equiv x/L, \nonumber\\
t ^\ast & \equiv t/\tau_{A}, \nonumber\\
\rho^\ast( t^\ast ) & \equiv \rho(t)/\rho_0 ,\nonumber\\
p^\ast (x^\ast ,t^\ast)  & \equiv p(x,t)/p_0,\nonumber\\
B_y^\ast (x^\ast, t^\ast)  & \equiv B_y(x,t)/B_0, \nonumber\\
v^\ast (x^\ast, t^\ast)  & \equiv v_x(x,t)/V_A,\nonumber
\end{align}
where $L$ is the initial current sheet thickness, $V_A$ is the initial Alfven speed, $\tau_A \equiv L/V_A$ is an Alfven timescale and 
the initial values of the physical quantities are denoted by 0.
The dimensionless equations can be written as 
\begin{align}
\frac{\partial}{\partial t^\ast} \rho^\ast & =  -\frac{\partial}{\partial x^\ast}\left( \rho^\ast v^\ast \right) , \label{1} \\
\frac{\partial}{\partial t^\ast} \left( \rho^\ast v^\ast\right) & =  -\beta_0 \frac{\partial}{\partial x^\ast}p^\ast 
	- \frac{\partial}{\partial x^\ast} \left( \rho^\ast v^{\ast 2} \right) 
	-\frac{\partial}{\partial x^\ast}\left(\frac{B_y^{\ast 2}}{2} \right), \label{2}\\
\frac{\partial}{\partial t^\ast} B_y^\ast & =  -\frac{\partial}{\partial x^\ast}\left( v^\ast B_y^\ast \right),\label{3}
\end{align}
where $\beta_0 = p_0/(B_0^2/2)$ is the initial plasma beta.

We introduce a scale factor $a(t)$ as follows: 
\begin{align}
v^\ast = \frac{\dot{a}(t^\ast)}{a(t^\ast)}x^\ast,  \label{4}
\end{align}
where a dot represents the time derivative. 
An Ansatz is that the velocity is linear in $x^\ast$.
For simplicity, we assume that $\rho^\ast$, the density in the current sheet, is spatial uniform.
From equation~(\ref{1}) and~(\ref{4}), we obtain
\begin{align}
\rho^\ast = \frac{1}{a(t^\ast)}.\label{5}
\end{align}
We also assume that $B_y^\ast$ is linear in $x^\ast$; $B_y^\ast(x^\ast,t^\ast) = B^\ast(t^\ast) x^\ast$.
Then from equation~(\ref{3}) and~(\ref{4}), we obtain 
\begin{align}
B^\ast = \frac{1}{a(t^\ast )^2}.\label{6}
\end{align}
We take the following form of the scaling factor, 
\begin{align}
a(t^\ast) \propto \left( \frac{t_0-t}{t_0} \right) ^k = \left( 1-\frac{t^\ast}{t_0^\ast} \right) ^k , \label{7}
\end{align}
where $t_0$ is the explosion time when the thickness of the current sheet goes to 0.
Neglecting the gas pressure term in (\ref{2}), we obtain the index $k$ and explosion time $t_0$ from equation(\ref{2}) and (\ref{7}):
\begin{align}
a(t^\ast) & \propto \left( 1-\frac{t^\ast}{t_0^\ast} \right)^{2/3}, \label{8} \\
t_0^\ast & = \frac{\sqrt{2}}{3}.
\end{align}
From equation~(\ref{8}), we obtain the following expressions:
\begin{align}
v^\ast (x^\ast ,t^\ast ) & \propto  \left( 1-\frac{t^\ast}{t_0^\ast} \right)^{-1} x^\ast, \label{9}\\
\rho^\ast (t^\ast) & \propto  \left(1-\frac{t^\ast}{t_0^\ast} \right)^{-2/3}, \label{10}\\
B_y^\ast (x^\ast ,t^\ast) & \propto  \left(1-\frac{t^\ast}{t_0^\ast}\right)^{-4/3} x^\ast \label{11}.
\end{align}
Note that these scaling laws are the same as those original obtained by \citet{imshennik67}.

If the plasma adiabatically evolves, we obtain the gas pressure in the following form, 
\begin{align}
p^\ast (t^\ast ) \propto \left( 1-\frac{t^\ast}{t_0^\ast} \right)^{-10/9}\label{12}.
\end{align}
This indicates that the gas pressure at the center should explode as $t \rightarrow t_0$.
Therefore the pinching by a Lorentz force must stop, which cannot be described by the linear theory.
To study the evolution near and after the explosion time, we numerically investigated the implosion of a current sheet using 1D MHD simulations.

\section{Basic Equations and Initial Conditions of Numerical Simulations}
All the physical quantities are functions of $x$ and $t$, and a magnetic field is considered only in the $y$-direction ($B_y$).
The basic equations are as follows:
\begin{align}
\frac{\partial \rho}{\partial t}&+\frac{\partial}{\partial x} \left( \rho v_x \right) = 0,\label{13}\\
\frac{\partial}{\partial t} \left( \rho v_x \right)&+ \frac{\partial}{\partial x} \left( \rho {v_x}^2 + p - \frac{{B_y}^2}{8\pi}\right)= 0,\label{14}\\
\frac{\partial B_y}{\partial t} &+\frac{\partial}{\partial x} \left( v_y B_x -B_y v_x \right) = 0,\label{15}\\
\frac{\partial e}{\partial t}&+\frac{\partial}{\partial x} \left[ \left( e+p + \frac{B_y^2}{8\pi}\right) v_x\right] = 0\label{16}, 
\end{align}
where the total energy density $e$ is defined as
\begin{align}
e & \equiv  \frac{p}{\gamma - 1} + \frac{\rho {v_x}^2}{2} + \frac{{B_y}^2}{8\pi}.\label{17}
\end{align}

We adopted the self-similar solution of \citet{tajima87} as the initial condition of the numerical simulations.
The initial condition is not an MHD equilibrium, and after the simulation's start, the current sheet is pinched by a Lorentz force as shown in Figure~\ref{fig11}. 
The calculation domain is in the range of $0 \leq x/L \leq 2$, where $L$ is the initial width of the current sheet.
An anti-parallel magnetic field is given by
\begin{align}
B_y(x) = B_0{\rm tanh}\left( \frac{x}{L} \right)\label{19}, 
\end{align}
where $B_0$ is defined as $\mid B_y(\pm L) \mid$.
For simplicity, we set the initial gas pressure $p$ and the initial electron density $\rho$ to be spatially uniform.
The initial gas pressure is normalized by the initial magnetic pressure ${B_0}^2/8\pi$.
The initial velocity field is given by
\begin{align}
v_x(x) = v_0{\rm tanh}\left( \frac{x}{L} \right)\label{20}. 
\end{align}
In our numerical simulations, the total grid number is fixed to 40000, which is adequately larger than 512 in \citet{forbes82}.
The numerical scheme we adopted is based on the HLLD scheme, which is a fully shock-capturing scheme \citep{miyoshi05}.

\section{Results of Numerical Simulations}
\subsection{Time Evolution of a Current Sheet in Numerical Simulation}
\begin{figure}
	\begin{center}
		\includegraphics[scale=0.90]{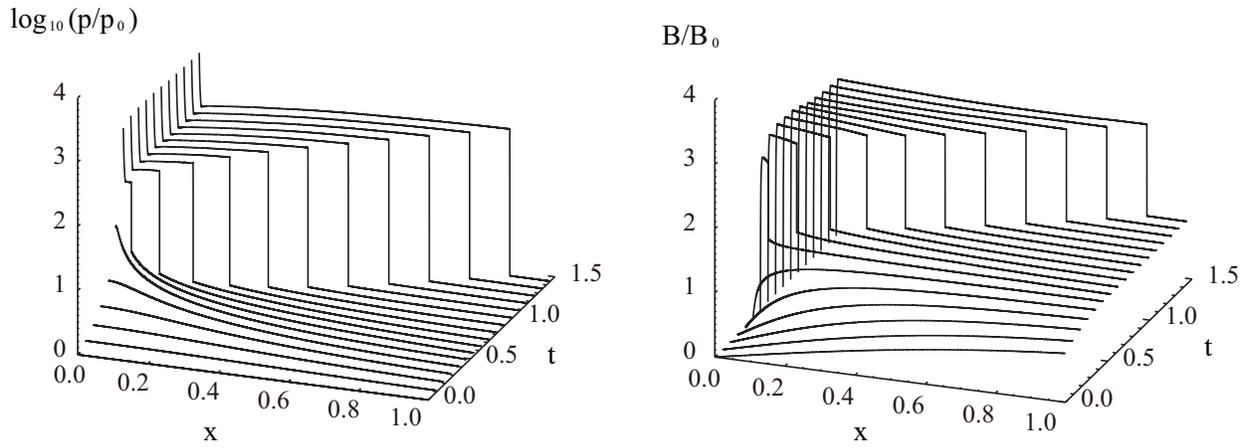}
		\caption{The time evolution of the implosion process of $p$ and $B_y$ with $\beta_0 = 10^{-2}$.
			The time interval and the last time are respectively $\Delta t = 0.9$ and $t_{\rm end} = 1.5$ in the unit of the Alfven time.}
		\label{fig15}
	\end{center}	
\end{figure}

\begin{figure}
\begin{center}
\includegraphics[scale=.90,clip]{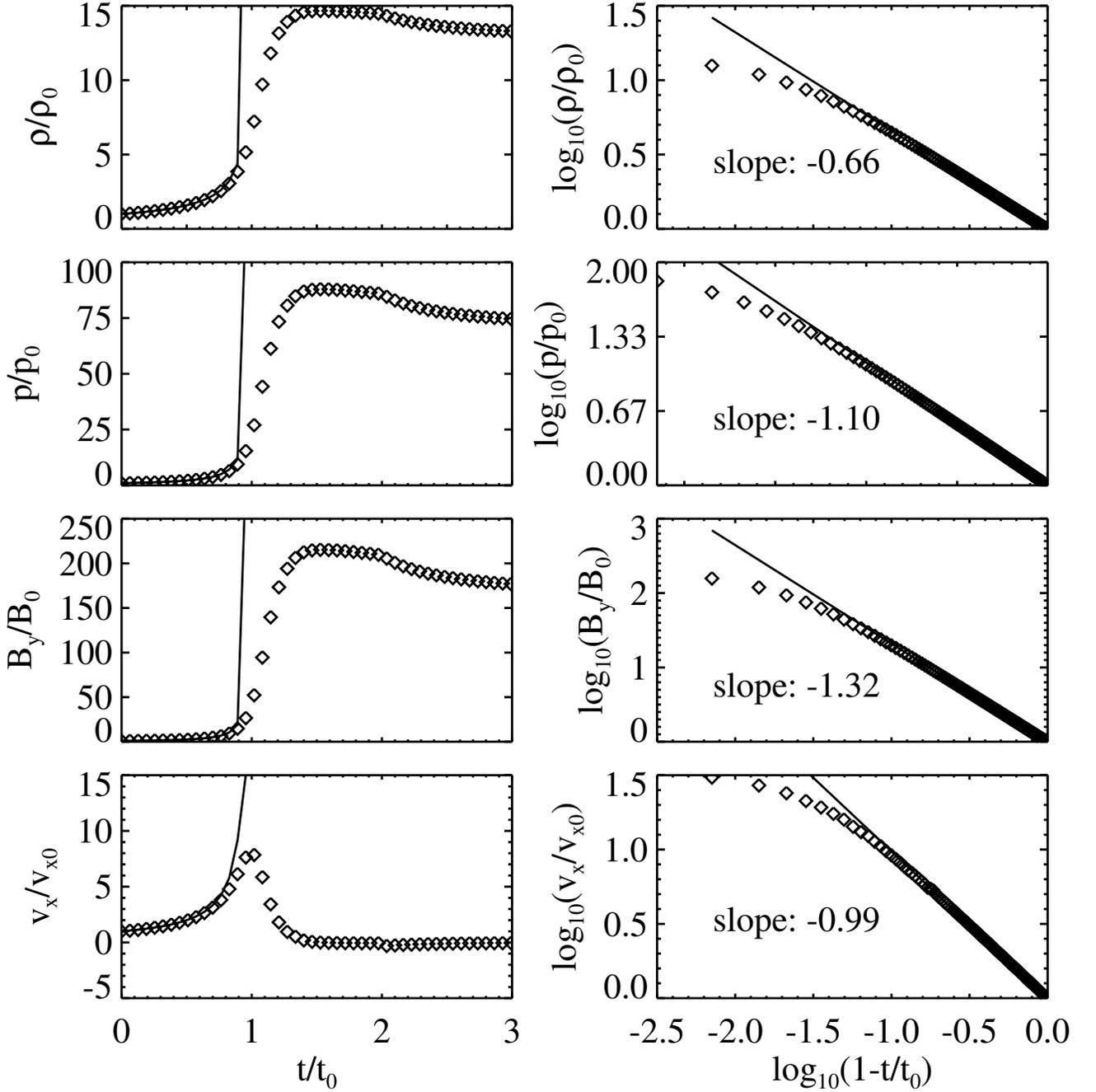}
\caption{Time evolution of $\rho, p, B_y$ and $v_x$ near the center of the current sheet with $\beta_0 = 10^{-2}$.
The initial values are denoted by subscript 0.
In the left column,  the diamonds represent the numerical results 
and the solid lines represents the analytical self-similar solution proposed in \citet{tajima87}.
In the right column, he diamonds represent the numerical results and the solid lines represents the fitting lines by the least squares method.
The indexes of the result of fitting are quite similar to them of the self-similar solution proposed in \citet{tajima87}.}
\label{fig1}
\end{center}
\end{figure}

As a typical case, here we show the time evolution of the case with $\beta _0 = 10^{-2}$ (see Figure~\ref{fig15}).
During $t < t_0$, the self-similar evolution predicted in \citet{tajima87} is confirmed with plasma gas pressure (see Figure~\ref{fig1}).
At $t \sim t_0$, the implosion leads to the formation of a shock (MHD fast-mode shock) as a result of a significant increase in the gas pressure at the center of the current sheet.
As shown in Figure~\ref{fig15}, the compressed plasma is continuously left behind the shock as the shock propagates outward.
Note that $p/\rho ^\gamma$ is almost kept constant at the center of the current sheet, which indicates that the physical quantities there adiabatically evolve.
To investigate the implosion process with the gas pressure, we performed the numerical simulations with various values of initial gas pressure and the spatial constant value of initial velocity.
As a result of numerical simulations, we found the power law behavior of physical quantities in the pinched current sheet behind the shock.
As shown in next section, we were able to analytically derive the scaling law with respect to $\beta_0$.

\subsection{$\beta_0$ Dependence of Physical Quantities in Current Sheet}
\begin{figure}
\begin{center}
\includegraphics[scale=.90]{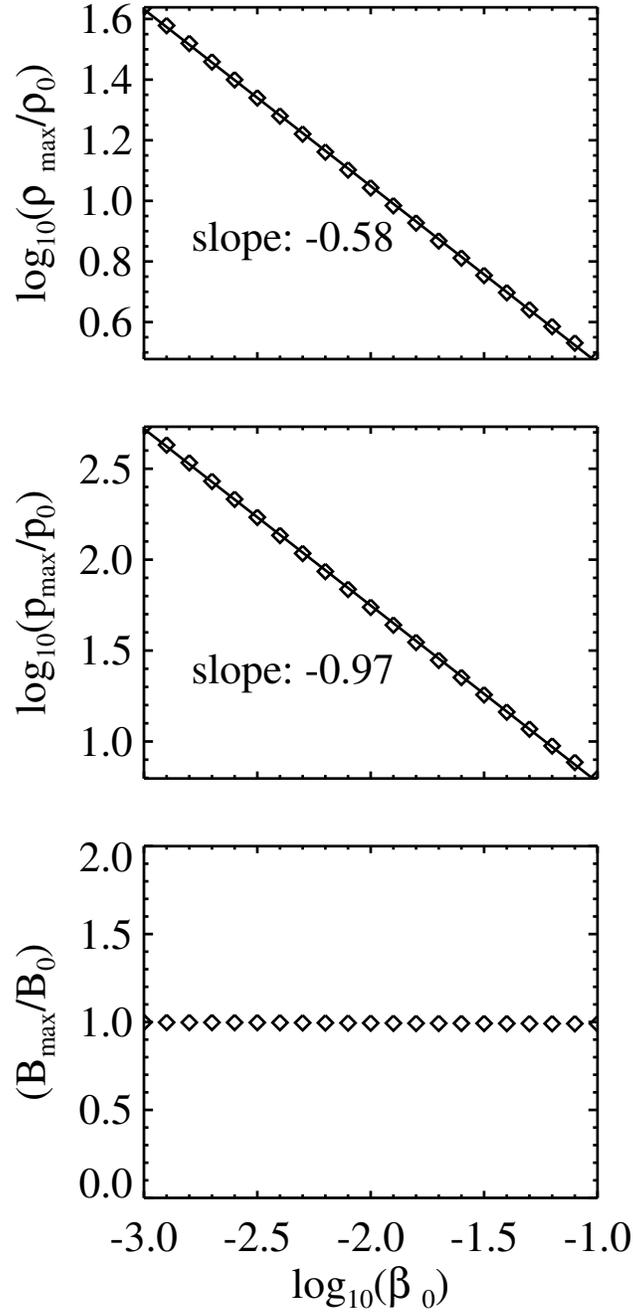}
\caption{The $\beta_0$ dependence of $\rho_{\rm max},p_{\rm max}$ and $B_{\rm max}$.
The diamonds represent the result of the numerical simulations and the solid lines represent the fitting lines by the least squares method.}
\label{fig3}
\end{center}
\end{figure}

We investigate the $\beta _0$ dependence of the physical quantities in the current sheet.
The range of $\beta _0$ is $-3 \leq \log \beta _0 \leq -1$.
The $\beta_0$ dependence of the density, pressure, and magnetic field strength just behind the shock is shown in Figure~\ref{fig3}.
Figure~\ref{fig3} implies the following power law relations:
\begin{align}
\rho /\rho_0 & \propto \beta_0^{-0.58}, \label{51} \\
p/p_0 & \propto \beta_0^{-0.97}, \label{52}\\
B_y/B_{y0} & \sim {\rm const}. \label{53} 
\end{align}

\begin{figure}
\begin{center}
\includegraphics[scale=.90]{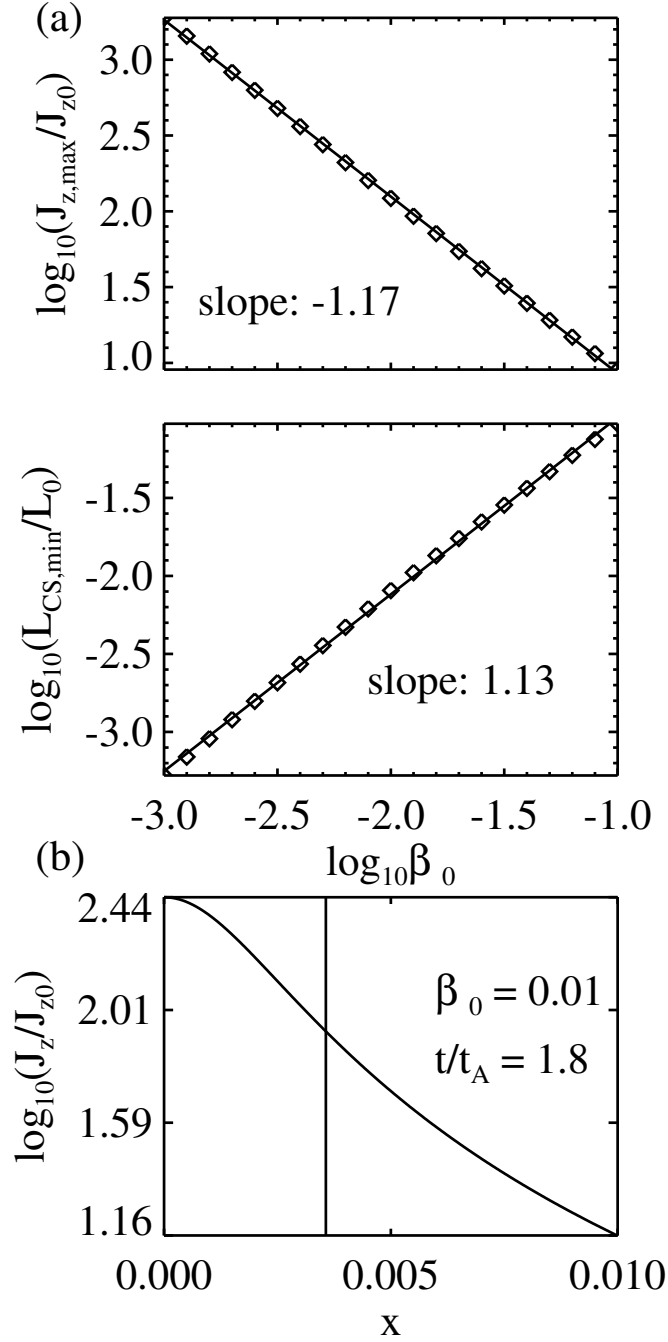}
\caption{(a)The $\beta_0$ dependence of the maximum current density $J_{\rm max}$(top) and of the minimum thickness of the current sheet $L_{\rm min}$(bottom).
The diamonds represent the result of the numerical simulations and the solid lines represent the fitting lines by the least squares method.
(b) The current density distribution with $\beta_0 = 0.01$ and $t/t_{\rm A} = 1.8$, where $t_{\rm A}$ is the Alfven time.
The vertical line indicates the current sheet thickness, $L_{CS} \equiv B_{\rm max}/J_{\rm max}$.}
\label{fig4}
\end{center}
\end{figure}

We also investigated the $\beta _0$ dependence of the thickness of a current sheet, $L_{\rm min}$.
We define $L_{\rm min}$ as follows:
\begin{align}
L(t) = \frac{B_{\rm max}(t)}{J_{\rm max}(t)}, \label{37}
\end{align}
where $B_{\rm max}(t)$ and $J_{\rm max}(t)$ are respectively 
\begin{align}
B_{\rm max} & \equiv \max \left( B_y(x,t) \right) , \nonumber \\
J_{\rm max} & \equiv \max \left( J_z(x,t) \right) . \nonumber
\end{align}.
As a result of the numerical simulations, we found a scaling law of $L_{\rm min}$ (see Figure~{\ref{fig4}}) as follows:
\begin{align}
L_{\rm min} \propto {\beta _0}^{1.177}. \label{38}
\end{align} 

\begin{figure}
\begin{center}
\includegraphics[scale=.90]{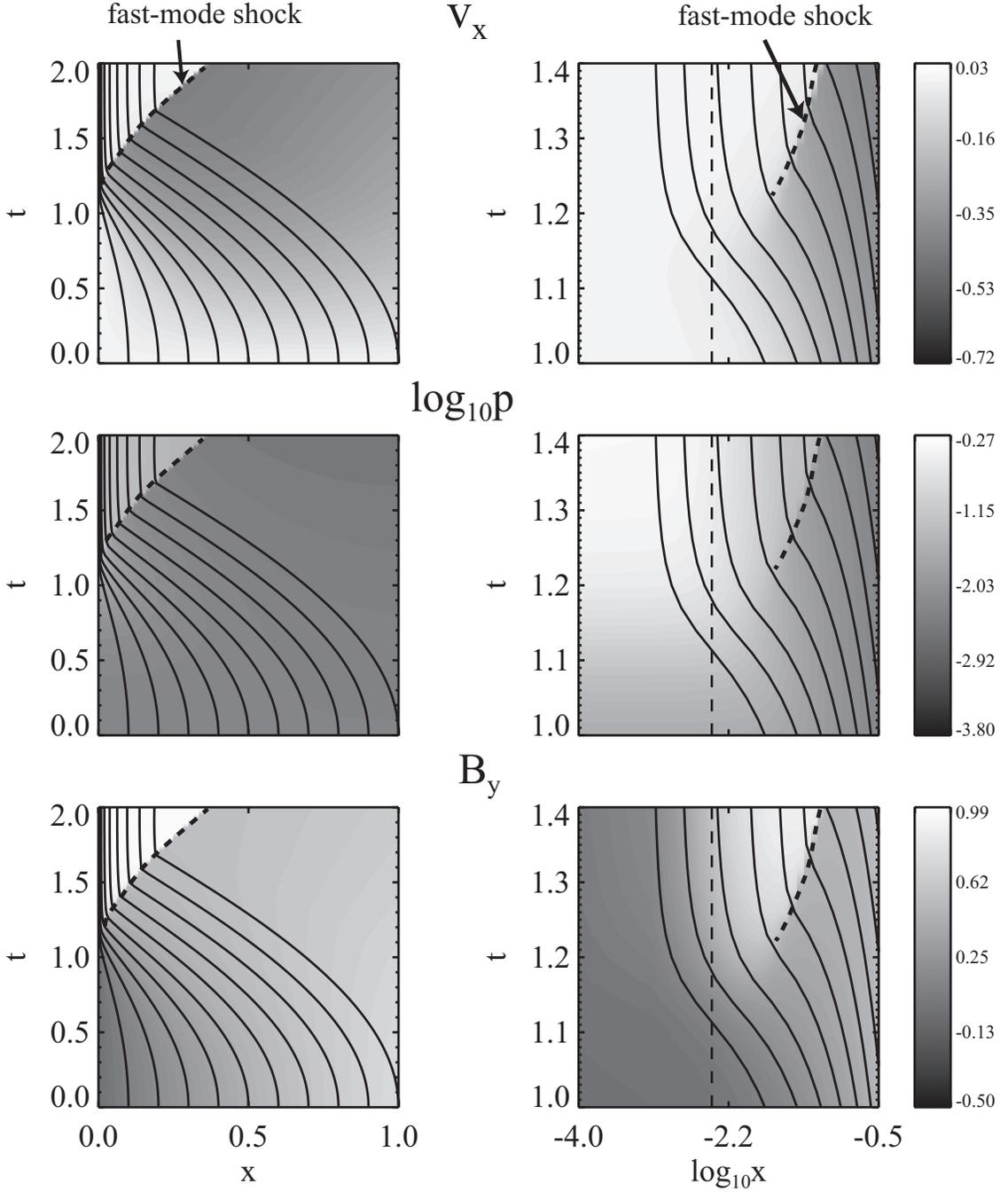}
\caption{Time-distance diagrams of $v_x$ (top), $\log_{10}p$(middle) and $B_y$(bottom).
The color contour denotes the values of the quantities.
The solid lines represent the trajectories of the selected Lagrangian particles.
The vertical dashed lines represent the current sheet thickness, $L_{CS} \equiv B_{\rm max}/J_{\rm max}$. 
The left column shows region of  $0 \leq x \leq 1$ and $0 \leq t \leq 2$. 
The right column shows the limited region of $-4 \leq \log_{10}x \leq -0.5$ and $1.0 \leq t \leq 1.4$.
}
\label{fig6}
\end{center}
\end{figure}

To study the formation of the pinched current sheet in detail, we tracked selected Lagrangian particles that are initially located in the initial current sheet.
 Figure~{\ref{fig6}} shows the trajectories of the Lagrangian particles on the time-distance diagrams of $v_x, \log_{10}p$ and $B_y$.
 From the Figure~{\ref{fig6}}, we found that the pinched current sheet is formed from a part of the initial current sheet, not from the whole of it.

\section{The Analytical Discussion about the $\beta_0$ Dependence of Physical Quantities}
In this section, we aim to derive the $\beta_0$ dependence of the physical quantities in the pinched current sheet.
As shown in $\S 4$, we confirmed the pinching process by the Lorentz force and the propagation of the MHD fast-mode shocks, as shown in \citet{forbes82}.
Behind the fast-mode shock, the physical quantities are determined by the Rankine-Hugoniot equations:
\begin{align}
\frac{v_{x2}}{v_{x1}} & =  \frac{1}{X}, \label{28}\\
\frac{B_{y2}}{B_{y1}} & =  X, \label{29}\\
\frac{p_2}{p_1} & =  \gamma {\bar{M_1}}^2 \left( 1-\frac{1}{X} \right) + \frac{1-X^2}{\beta _1}+1, \label{30}
\end{align}
where the physical quantities in front of and behind the shock are respectively denoted by 1 and 2, $\bar{M_1} = v_{x1}/c_{s1}$ is the Mach number, and $X = \rho_2/\rho_1$ is the density ratio, which is the positive solution of
\begin{align}
2\left( 2- \gamma \right) X^2 + \left[ 2\beta _1 + \left( \gamma -1 \right)\beta _1{\bar{M_1}}^2 + 2 \right] \gamma X \nonumber \\
- \gamma \left( \gamma +1 \right) \beta _1{\bar{M_1}}^2 = 0. \label{31}
\end{align}   
From equation~(\ref{29}) and~(\ref{31}), the compression ratio is limited in the range  of
\begin{align}
1 < \frac{B_2}{B_1} <\frac{\gamma + 1}{\gamma -1} , \label{32}
\end{align}
where the upper limit is 4 for $\gamma = 5/3$.
Therefore the maximum value of the strength of the magnetic field should weakly depend on $\beta_0$ ($B_{\rm max}/B_0 \sim \rm{const} $ with respect to $\beta_0$, equation(\ref{53})).
The current sheet, which is pinched by the shocks, is in the MHD equilibrium (i.e. $v_x = 0$).
This means that the gas pressure at the center of the current sheet, $p_{\rm max}$, should be comparable to the magnetic pressure outside of the current sheet:
\begin{align}
p_{\rm max} & \sim \frac{{B_{\rm max}}^2}{8\pi} \sim \frac{{B_0}^2}{8\pi}, 
\end{align}
which gives the relation(\ref{52})
\begin{align}
\frac{p_{\rm max}}{p_0} & \sim  {\beta _0}^{-1}, \label{34}
\end{align}
where $p_0$ is the initial gas pressure.
From equation~(\ref{34}) and the adiabatic condition, the density at the center (which takes the maximum value, equation(\ref{51})) can be written by
\begin{align}
\frac{\rho _{\rm max}}{\rho _0} & \sim  \left( \frac{p_{\rm max}}{p_0} \right) ^{1/\gamma}, \nonumber\\
			     & \sim  {\beta _0}^{-3/5}, \label{35}
\end{align}
where $\rho _0$ is the initial plasma density and $\gamma = 3/5$ is the adiabatic index. 

To derive the $\beta_0$ dependence of $L_{\rm min}$, the relation~(\ref{38}), we defined $B_1$, the magnetic field strength of the maximum value in the pinched current sheet.
As shown in Lagrangian particle trajectories, note that the pinched current sheet is formed from not the whole of the initial current but a part of it.  
We also defined the length of the part in the initial current sheet as $L_0^\ast$ and the maximum value of the magnetic field strength in the part as $B_0^\ast$.
In our numerical simulations, we confirmed $L_0^\ast \ll L_0$ by the Lagrangian particle motions.
Near the center of the initial current sheet, the magnetic field strength linearly depend on $x$,
\begin{align}
\frac{L_0^\ast}{L_0} \sim \frac{B_0^\ast}{B_0}. \label{39}
\end{align}
We considered the conservation law of a magnetic flux,
\begin{align}
B_0^\ast L_0^\ast & = B_1L_{\rm min}, \nonumber\\
	& = B_0 L_{\rm min}. \label{40}
\end{align}
From equation~(\ref{39}) and~(\ref{40}), we obtained
\begin{align}
\frac{L_{\rm min}}{L_0} \sim \left( \frac{L_0^\ast}{L_0} \right)^2 . \label{41}
\end{align}
Finally, from the mass conservation law, 
\begin{align}
\frac{L_0}{L_{\rm min}} = \frac{\rho_{\rm max}}{\rho _0} \sim \beta _0^{-3/5}. \label{43}
\end{align}
From equation~(\ref{41}) and~(\ref{43}), we derived
\begin{align}
\frac{L_{\rm min}}{L_0} \sim \beta _0^{6/5}, \label{44}
\end{align}
which is consistent with equation~(\ref{38}).

\section{Discussion}
In this paper, we studied the implosion process of current sheets in a low-$\beta$ plasma using one-dimensional ideal MHD simulations.
We confirmed that the self-similar solution by \citet{tajima87} holds before the explosion time.
The plasma adiabatically evolves till the explosion time.
At $t \sim t_0$, the behavior of the current sheet deviates from the analytical self-similar solution and the MHD fast-mode shocks 
are formed because of the increase in the gas pressure at the center of the current sheet.
After a pair of MHD fast-mode shocks propagate away from the current sheet, the pinched current sheet is formed between a pair of the shocks.
We studied the $\beta_{0}$ dependence of the maximum values of the physical quantities inside the current sheet and the minimum thickness of the current sheet (equation~(\ref{34}), (\ref{35}) and (\ref{44})).

Contrast to our study, the $\eta$ dependence of the physical quantities and of the reconnection rate were investigated by \citet{mcclymont96} in MHD simulations similar to ours, where $\eta$ is the magnetic diffusivity.
They found that the effect of the gas pressure does not significantly change the dependence of the thickness of the current sheet on $\eta$,
which means that our scaling laws can be easily extended to the cases with the resistivity.   

In \citet{mclaughlin09}, the collapse of a null point to a current sheet triggered by MHD fast-mode waves was investigated.
In their simulations, successive current sheet pinching was observed.
The pinching is done by the gas pressure of the heated plasma in the outflow region, while the pinching in our study is done by a Lorentz force.
To extend this study, we are currently performing two-dimensional MHD simulations of the coalescence process of plasmoids and studying the pinching process of the current sheet formed between plasmoids in detail.

\begin{figure}
\begin{center}
\includegraphics[scale=.90]{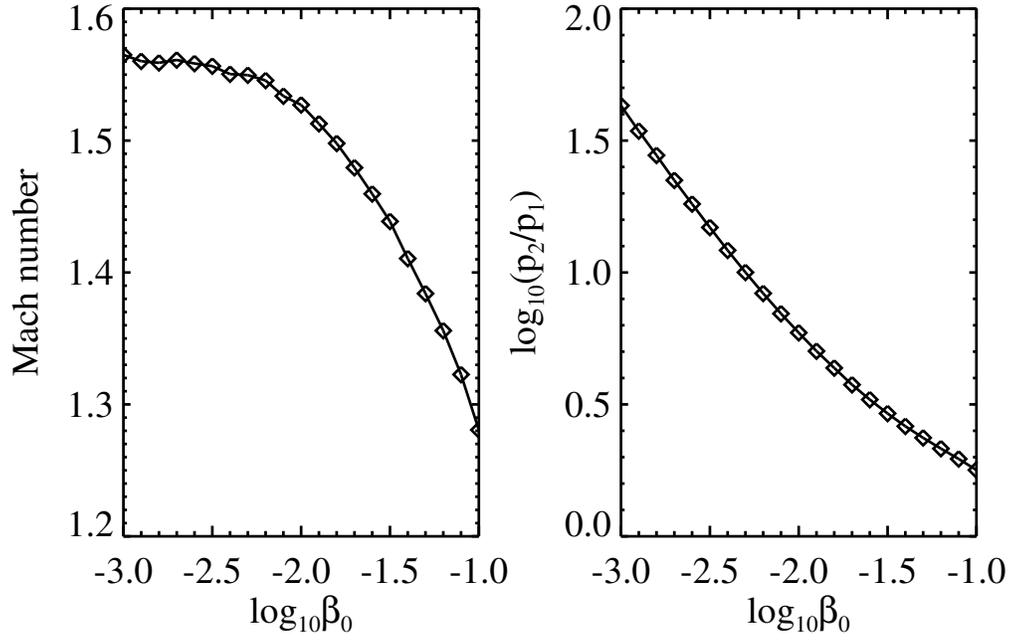}
\caption{The $\beta_0$ dependence of the fast-mode Mach number and $p_2/p_1$, where the gas pressures behind and in front of the
 shock are respectively denoted by the subscript 1 and 2.
}
\label{fig20}
\end{center}
\end{figure}

\begin{figure}
\begin{center}
\includegraphics[scale=.90]{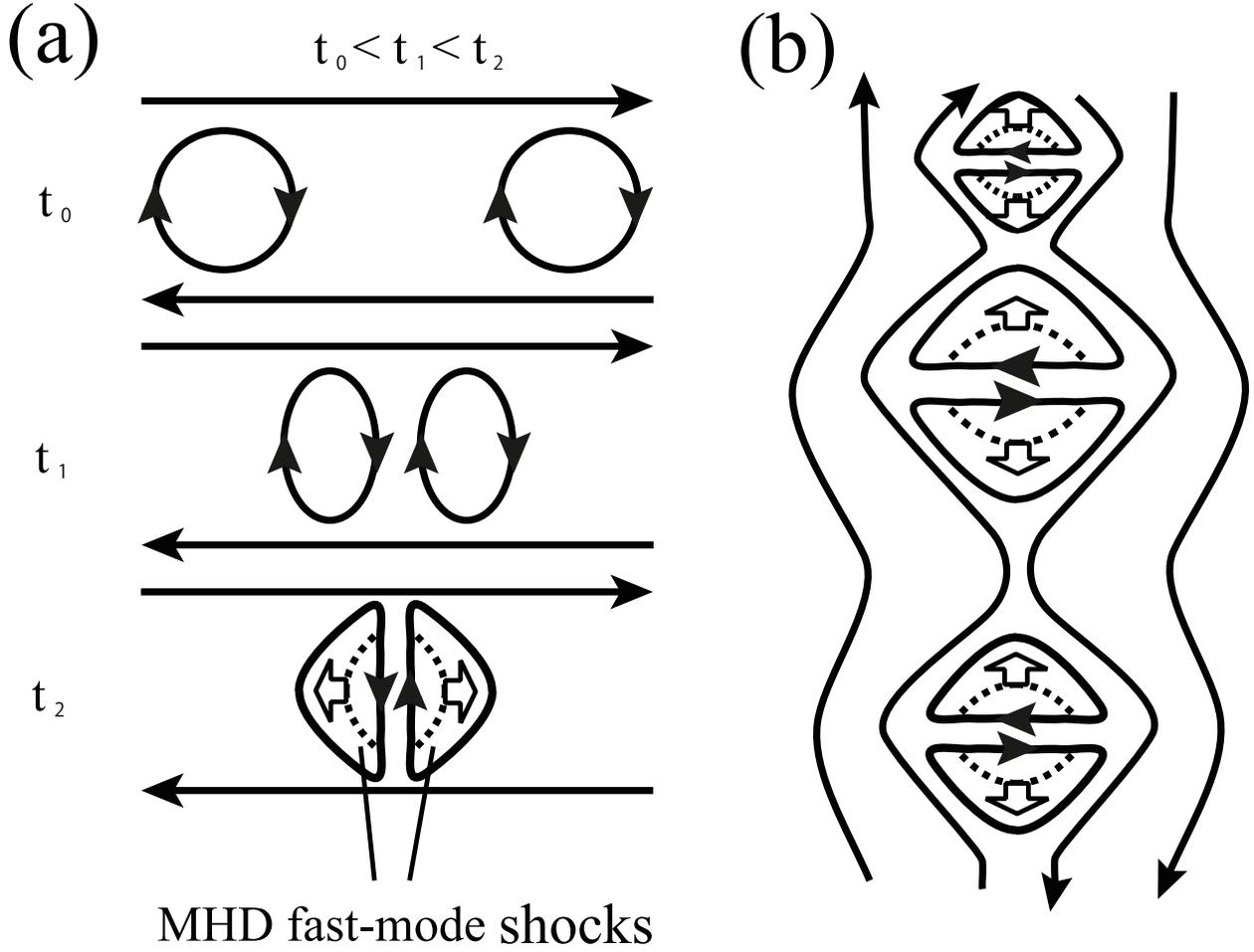}
\caption{(a)A schematic picture of the time evolution of the pinching process and the formation of MHD fast-mode shocks. 
(b)A schematic picture of our conjecture about an actual high-Lundquist number reconnection process, where particles could be 
accelerated at fast-mode shocks generated by the coalescence of plasmoids.  
}
\label{fig22}
\end{center}
\end{figure}

We analyzed the $\beta_0$ dependence of the fast-mode Mach number and the compression ratio, $p_2/p_1$(see Figure~\ref{fig20}).
Figure~\ref{fig20} shows that the fast-mode Mach number, $\bar{M_1}$, weakly depend on $\beta_0$.
We also found that the compression ratio of the density and magnetic field, $X$, weakly depend on $\beta_0$.
Therefore, from equation~(\ref{30}), we obtain
\begin{align}
\log_{10} \left( \frac{p_2}{p_1} \right) \propto \beta_1 \sim \beta_0 . \label{56}
\end{align}
This $\beta_0$ dependence of the compression ratio, $p_2/p_1$, is similar to equation(\ref{56}) with $\beta_0 < 10^{-2}$.
In recent studies, it is indicated that particles can be accelerated in contracting plasmoid \citep{drake06} and in plasmoids crossing fast-mode shocks\citep{nishizuka13}.
Figure~\ref{fig22}(a) shows an overview of the implosion process of the current sheet between plasmoids and the formation of the MHD fast-mode shocks.
In our study, we found that fast-mode shocks can be formed during the implosion process in low-$\beta$ plasmas. 
Applying the shock formation in the implosion process to a coalescence process, we conjecture that particles in the plasmoids are efficiently accelerated 
at shocks generated by the coalescence of plasmoids through a Fermi acceleration process (see Figure~\ref{fig22} (b)).

We investigated the parameter region where our scaling laws are valid.
Since we assume the framework of non-relativistic ideal MHD, (1) the drift velocity of the electrons needs to be sufficiently smaller than the light speed and (2) the thickness of the current sheet needs to be larger than the ion skin depth.
The electron drift velocity can be written as follows: 
\begin{align}
J & = nev_{{\rm drift},e}, \nonumber\\
	& = \frac{c}{4\pi}\nabla \times \mbox{\boldmath $B$} \sim \frac{c}{4\pi} \frac{B}{L}, \label{45}
\end{align}
where $L$ is the thickness of current sheet.
From equation~(\ref{45}), $v_{{\rm drift},e}$, the drift velocity of electrons, is given as
\begin{align}
	\frac{v_{{\rm drift},e}}{c} \sim \frac{1}{4\pi e} \frac{B}{nL}. \label{46}
\end{align}
Applying the scaling laws of the relation (\ref{34}), (\ref{35}) and (\ref{44}) to equation~(\ref{46}), we obtain
\begin{align}
	\frac{v_{{\rm drift},e}}{c} &\sim \nonumber\\
 &10^{-4} \left( \frac{\beta_0}{\beta_{-2}}\right)^{-3/5} \left( \frac{B_0}{B_{1}} \right) \left(\frac{n_0}{n_{9}}\right)^{-1} \left( \frac{L_0}{L_{6}} \right)^{-1}, \label{47}
\end{align}
where $\beta_{-2} = 10^{-2}, B_1 = 10{\rm G}, n_9 = 10^9{\rm cm^{-3}}$ and $L_6 = 10^6{\rm cm}$.
Since we study non-relativistic MHD processes, $v_{{\rm drift},e}/c < 1$ is required.
This leads to 
\begin{align}
	\log_{10}\left(\frac{L_0}{L_{6}}\right)  > & - \log_{10}\left(\frac{n_0}{n_{9}}\right) -\frac{3}{5}\log_{10}\left(\frac{\beta_0}{\beta_{-2}}\right)\nonumber\\
 &+ \log_{10}\left(\frac{B_0}{B_{1}}\right) - 4. \label{48}
\end{align}

Since the thickness of the current sheet must be larger than the ion skin depth, we similarly obtain 
\begin{align}
	\frac{\lambda _i}{d} & = \frac{1}{c} \sqrt{\frac{4\pi e^2}{m_i}} \frac{\sqrt{n}}{d}, \nonumber\\
		& \sim 10^{-4} \left(\frac{\beta_0}{\beta_{-2}}\right)^{-3/2} \left(\frac{n_0}{n_{9}}\right)^{1/2}\left(\frac{L_0}{L_{6}}\right)^{-1} < 1, \label{49}
\end{align}
where $\lambda _i$ is the ion skin depth.
We can transform equation(\ref{49}) as
\begin{align}
	\log_{10}\left(\frac{L_0}{L_{6}}\right) > &\frac{1}{2}\log_{10}\left(\frac{n_0}{n_{9}}\right) \nonumber\\
	&- \frac{3}{2}\log_{10}\left(\frac{\beta_0}{\beta_{-2}}\right) -4. \label{50}
\end{align}

\begin{figure}
\begin{center}
\includegraphics[scale=.90]{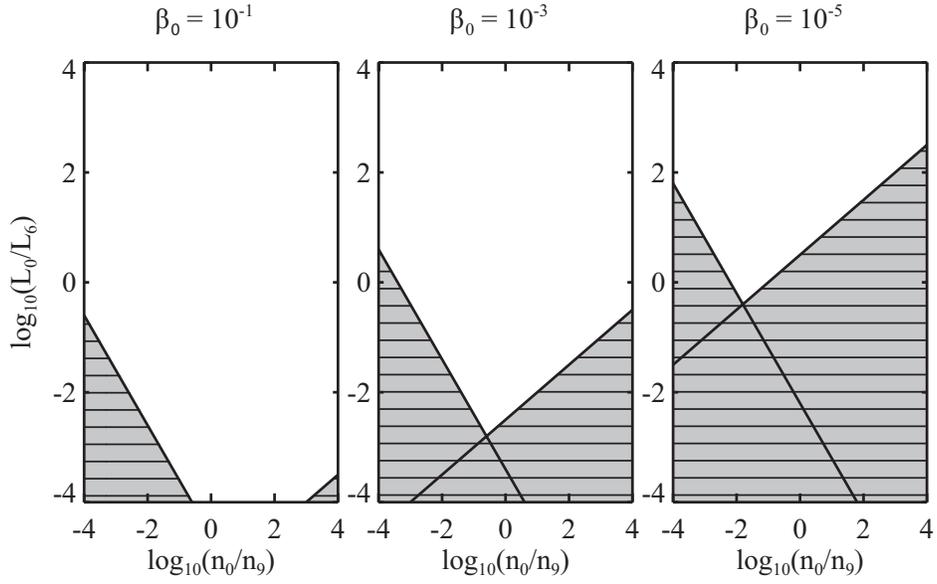}
\caption{
The parameter regions where both inequality~(\ref{48}) and~(\ref{50}) are valid (i.e. the non-relativistic MHD approximation is valid).
The cases with $\beta_0 = 10^{-1}$(left), $\beta_0 = 10^{-3}$(middle) and $\beta_0 = 10^{-5}$(right) are shown
, where $n_9 = 10^9{\rm cm^{-3}}$ and $L_6 = 10^6{\rm cm}$.
The shaded zones represent the parameter regions where our scaling laws do not hold.
}
\label{fig13}
\end{center}
\end{figure}

In Figure~\ref{fig13}, we show the parameter region using inequality~(\ref{48}) and (\ref{50}) with the value in the solar corona, $\beta_0 = 10^{-2}$ and $B_0$ = $10$G.
It is found that our scaling laws are applicable in the solar corona.
These scaling laws will help us to estimate physical quantities in a current sheet formed in a low-$\beta$ plasma.
Describing the evolution of the pinching process in an extremely low-$\beta$ plasma like the atmosphere of the neutron stars will be our future work.

\acknowledgments

We thank Dr. Jin Matsumoto for fruitful discussions and comments. 
Shinsuke Takasao acknowledges support by the Research Fellowship of the Japan Society for the Promotion of Science (JSPS).
This work was supported by the Grant-in-Aids from the Ministry of Education, Culture, Sports, Science and Technology of 
Japan (No. 25287039).

\clearpage

\end{document}